\documentclass[alpha-refs]{wiley-article}

\title{Towards hourly three-dimensional ensemble data assimilation of screen-level observations into coupled atmosphere-land models}

\author[123\authfn{1}]{Tobias Finn}
\author[4]{Gernot Geppert}
\author[15]{Felix Ament}

\affil[1]{Meteorological Institute, University of Hamburg, Hamburg, Germany}
\affil[2]{Meteorological Institute, University of Bonn, Bonn, Germany}
\affil[3]{International Max Planck Research School on Earth System Modelling, Max Planck Institute for Meteorology, Hamburg, Germany}
\affil[4]{Deutscher Wetterdienst, Offenbach, Germany}
\affil[5]{Max Planck Institute for Meteorology, Hamburg, Germany}

\corraddress{Tobias Finn, CEREA, École des Ponts and EDF R\&D, Île-de-France, France}
\corremail{tobias.finn@enpc.fr}

\presentadd[\authfn{1}]{CEREA, École des Ponts and EDF R\&D, Île-de-France, France}

\fundinginfo{Deutsche Forschungsgemeinschaft, Grant Number: 243358811}

\runningauthor{Finn et al.}

\begin{document}

\maketitle

\begin{abstract}
    We explore the potential of three-dimensional data assimilation for assimilating sparsely-distributed 2-metre temperature observations across the coupled atmosphere-land interface into the soil moisture.
    Using idealised twin experiments with the limited-area modelling platform TerrSysMP and synthetic observations, we avoid model biases and directly control errors in the initial conditions and observations.
    These experiments allow us to test hourly data assimilation with a localised ensemble Kalman filter, as often used for mesoscale data assimilation.
    We find here an error reduction of such an ensemble Kalman filter approach compared to daily-updating with a one-dimensional simplified extended Kalman filter.
    We attribute this improvement to the ensemble approximation of the sensitivities and the more frequent updates with the ensemble Kalman filter.
    The hourly updates result hereby into a positive assimilation impact during daytime and a neutral impact during night.
    With a three-dimensional ensemble Kalman filter, we can directly assimilate screen-level observations at their respective position into the soil moisture, skipping the otherwise needed spatial interpolation step.
    These findings suggest an emerging potential for the localised three-dimensional ensemble Kalman filter to hourly assimilate screen-level observations into coupled atmosphere-land models.
\end{abstract}

\section{Introduction}\label{introduction}
Driven by the incoming solar radiation, the sensible heat flux and evapotranspiration transfer energy and, thus, information from the land surface into the atmospheric boundary layer.
As a limiting factor for the evapotranspiration, the soil moisture controls also how much energy remains to establish a sensible heat flux.
Through this mechanism, the soil moisture modulates the temperature in the atmospheric boundary layer.
As a consequence, screen-level observations from the atmospheric boundary layer, such as the 2-metre temperature, encode information about conditions in the land surface like the soil moisture \citep{mahfouf1991, balsamo2007, liu2019}.

Simplified one-dimensional data assimilation methods make use of this physical relation in the operational cycle for weather forecast models \citep{ecmwf_ifs_da_2020, gomez2020, carrera2015}.
Here, the soil moisture supposedly influences the 2-metre temperature only at the same location in a single vertical column \citep{hess2001, rosnay2013}.
However, important processes in the land surface and atmospheric boundary layer act on scales that remain unresolved in current atmosphere-land models \citep{dirmeyer2017, orth2017, kauffeldt2015, best2015}.
The resulting biases lead to a negative impact of the assimilation of screen-level observations on the soil moisture analysis \citep{hess2001, drusch2007, draper2011, su2013, carrera2019, munoz-sabater2019}; in these settings, the land surface data assimilation acts as sink for model biases, correcting the soil moisture into arbitrary directions.

Recent developments in ensemble-based data assimilation \citep{carrassi2018} and fully-coupled atmosphere-land models for mesoscale weather prediction \citep{fatichi2016, vereecken2016} raise the question of how we can use these tools to extract the encoded information from screen-level observations, and how this can improve the land surface analysis.
To address these questions, we present idealised experiments designed to highlight the potential of a three-dimensional localised ensemble Kalman filter to assimilate sparsely-distributed 2-metre temperature observations into the soil moisture on an hourly basis.

\begin{figure}[ht]
    \centering
    \includegraphics[width=0.7\textwidth]{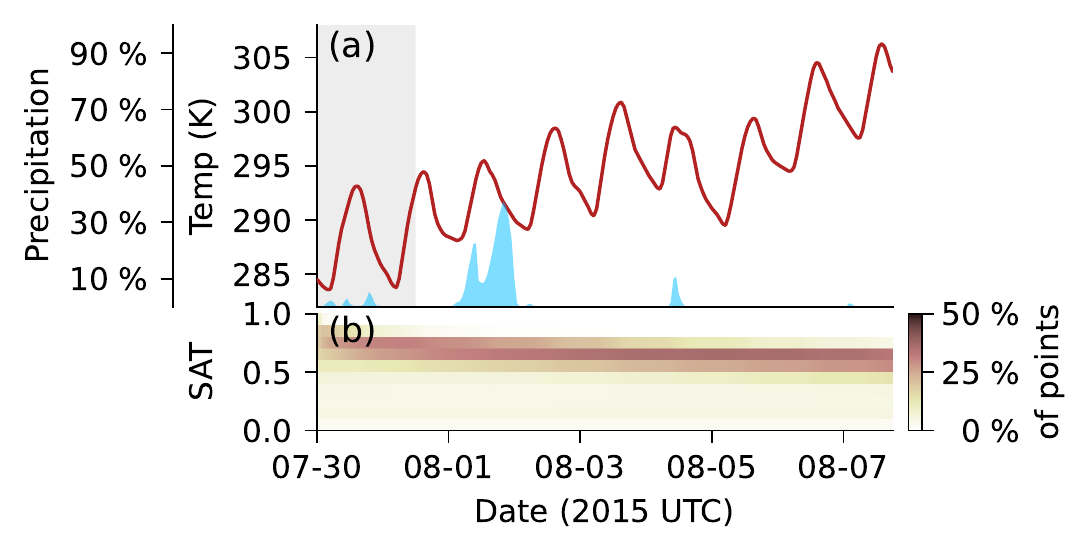}
    \caption{
        Weather overview extracted from the nature run with (a) the 10-metre temperature and the percentage of grid points with precipitation in blue, and (b) the soil moisture saturation as percentage of grid points within a $0.1$ bin. The grey-shaded period represents the spin-up time.\label{fig:overview}
    }
\end{figure}

To control errors and deactivate model biases, we employ idealised twin experiments.
Here, the same model configuration is used for data assimilation experiments as for a nature run.
This nature run acts as our reality, and we generate on its basis synthetic 2-metre temperature observations.
In addition, we use a Terrestrial System Modelling Platform \citep{shrestha2014, gasper2014} to model the interface between land surface and atmospheric boundary layer as accurately as possible.
Its process-based approach allows its use for regional mesoscale simulations \citep{gebler2017, kollet2018}.
As the selected simulation period features a mixture of soil-moisture-limited evapotranspiration regimes (see Figure \ref{fig:overview}), we expect a strong response of the 2-metre temperature to perturbations in the soil moisture.
We thus explore in this study the potential of three-dimensional ensemble Kalman filters for a strongly-coupled atmosphere-land interface in a favourable “perfect world” scenario.

Localised ensemble Kalman filters are state-of-the-art data assimilation methods, often used for mesoscale weather prediction \citep{gustafsson2018, schraff2016}.
These filters additionally show promising results for assimilation of satellite and in-situ observations of soil moisture into the land surface \citep{reichle2002, fairbairn2015, draper2019, bonan2020}.
Also, assimilating conventional and 2-metre humidity observations with an ensemble Kalman filter can improve the soil moisture analysis on a six hour interval \citep{lin2020}.
In our perfect world scenario, we confirm such a similarly emerging potential of hourly data assimilation of 2-metre temperature observations into the soil moisture analysis by comparing a ensemble Kalman filter to a one-dimensional simplified extended Kalman filter.
Using a three-dimensional ensemble Kalman filter, we gain a positive impact for assimilating screen-level observations at their respective position into the land surface, without an intermediate interpolation step, as needed for a simplified extended Kalman filter.
Therefore, we see this study as one of the first steps towards the use of three-dimensional ensemble Kalman filters for land surface data assimilation as used for atmospheric data assimilation.

\section{Experimental design}\label{experimental-design}
In this section, we shortly describe the used limited-area modelling system.
We additionally present the different concepts of the two data assimilation methods and the four performed experiments.

\subsection{Model}\label{model}
As coupled atmosphere-land model, we use the Terrestrial System Modelling Platform (TerrSysMP, \citealt{shrestha2014}; \citealt{gasper2014}).
In our configuration, TerrSysMP represents land surface processes by the column-based Community Land Model (CLM, Version 3.5, \citealt{oleson2004,olesonk.w.2008}).
OASIS3 \citep{valcke2013} couples CLM to the COnsortium for Small-scalle MOdelling model (COSMO, Version 4.21, \citealt{baldauf2011}).
For further information about the model components and their coupling, we refer to \citet{shrestha2014}.

Our model configuration resembles typical setups for mesoscale numerical weather prediction.
Here, COSMO has a horizontal resolution of 2.8 km and 50 full vertical levels, as used for the former operational COSMO-DE runs at the German Weather Service, DWD.
The horizontal resolution of CLM is set to 1 km with 10 vertical soil levels.
The simulations span a region of $267\,\text{km}\times302\,\text{km}$ in longitudinal and latitudinal direction, respectively, located in the south-western part of Germany with the embedded Neckar catchment.
As similarly used in \citet{schalge2021}, the heterogeneous land surface setup typifies conditions in mid-latitudes and Central Europe.
The lateral boundary conditions in the atmosphere are based upon one member of the COSMO-DE-EPS ensemble \citep{peralta2012} with a similar configuration to our atmospheric setup.
All our experiments have the same model configuration and lateral boundary conditions.

\subsection{Synthetic observations}\label{sec:observations}

To generate synthetic 2-metre temperature observations from the nature run, we use the diagnostic 2-metre temperature provided by COSMO.
In the ensemble Kalman filter, we hourly assimilate synthetic observations at the nearest 99 model grid points to actual observational sites.
For the simplified ensemble Kalman filter, we bilinearly interpolate the 2-metre temperature field from COSMO onto the CLM grid to provide observations at every CLM grid point once a day.
This interpolation is a simplification compared to operational applications, where often an intermediate optimal interpolation step maps sparse observations to the analysis grid \citep{rosnay2013, ecmwf_ifs_da_2020}.
We perturb all observations by independent errors drawn from a Gaussian distribution $\mathcal{N}(0, (\sigma^\text{o})^{2})$ with zero mean and observational error standard deviation $\sigma^\text{o} = 0.1\,\text{K}$.
With our observations, we assume in the data assimilation methods a perfect knowledge of the observation operator and its errors.

\subsection{Simplified extended Kalman filter}\label{simplified-extended-kalman-filter}

We use a simplified extended Kalman filter, SEKF, similar to the implementation proposed by \citet{rosnay2013} as our baseline data assimilation method.
With this method, we reflect a common approach of land surface data assimilation to update the soil moisture once a day at midnight based on screen-level observations at daytime \citep{balsamo2004, hess2001}.
The SEKF is applied to every CLM soil column independently to calculate a one-dimensional analysis at every grid point of a single, deterministic model run.
In case of spatially sparse observations, this involves an intermediate interpolation step to create an observation at every model grid point.
Using synthetic observations available at every grid point, $y^o$, simplifies this, and we assimilate the fully-observed 2-metre temperature field at 12:00 UTC (cf. section \ref{sec:observations}).

We update the soil moisture in the first seven layers of a model column, $\mathbf{x}^\mathrm{b}$.
They extend to a depth of $0.62\,\text{m}$; deeper layers barely contribute to evapotranspiration.
The background error covariance matrix is assumed to be constant and diagonal, $\mathbf{B}=(\sigma^\mathrm{b})^{2}\mathbf{I}$, with a uniform standard deviation $\sigma^\mathrm{b} = 0.01\,\text{m}^3 \text{m}^{-3}$ (as in \citealt{ecmwf_ifs_da_2020}).
The sensitivities of the 2-metre temperature to the soil moisture, $\mathbf{H}$, are approximated with finite differences.
To estimate the finite-differences approximation, one additional run for each updated soil layer is propagated from 00:00 UTC to 12:00 UTC with a perturbation of one standard deviation of the background covariance \citep{rosnay2013}.
The observational error covariance matrix is defined as diagonal based on the standard deviation of the observation error, $\mathbf{R}=(\sigma^\mathrm{o})^{2}\mathbf{I}$.
To derive observations of the background state, $H(\mathbf{x}^b)$, the diagnostic 2-metre temperature from COSMO is bilinearly interpolated to the CLM grid.
With these choices, the SEKF analysis for one soil column reads

\begin{align}
    \mathbf{x}^\mathrm{a} &= \mathbf{x}^\mathrm{b} + \mathbf{B}\mathbf{H}^\top(\mathbf{H}\mathbf{B}\mathbf{H}^\top + \mathbf{R})(y^\mathrm{o} - H(\mathbf{x}^\mathrm{b})).
\end{align}

\subsection{Localised ensemble transform Kalman filter}
\label{localised-three-dimensional-ensemble-kalman-filter}

Contrary to the SEKF, the localised ensemble transform Kalman filter \citep{bishop2001, hunt2007}, LETKF, is commonly used to calculate a three-dimensional analysis, for example in operational atmospheric data assimilation \citep{schraff2016, gustafsson2018}.
We use the LETKF with an ensemble of 40 members and update the soil moisture over the whole domain, $\left\{\mathbf{x}_i^\mathrm{b}\right\}_{i=1,\,...,\,40}$, from synthetic observations of 2-metre temperature at 99 grid points, $\mathbf{y}^\mathrm{o}$, every hour (cf. section \ref{sec:observations}).
The LETKF analysis is given as

\begin{align}
    \mathbf{x}_i^\mathrm{a} &= \sum_{j=1}^{40} w_{ij}\mathbf{x}_j^\mathrm{b},
\end{align}
where the weights $w_{ij}$ are calculated from the ensemble of the background state in observational space $H(\mathbf{x}_i^\mathrm{b})$ and the observation error covariance matrix, $\mathbf{R}=(\sigma^\mathrm{o})^{2}\mathbf{I}$.
The ensemble of observations $H(\mathbf{x}_i^\mathrm{b})$ is derived from the diagnostic 2-metre temperature from each of the 40 ensemble members at the same grid points that were used for the synthetic observations from the nature run.
The ensemble covariance of $\mathbf{x}_i^\mathrm{b}$ approximates the background error covariance, and the calculation of the weights $w_{ij}$ uses an implicit linearisation of the observation operator $H$ around the ensemble mean of the background state in observational space, $\overline{H(\mathbf{x}_i^\mathrm{b})}$ (cf. \citealt{hunt2007} for details).
Thus, the ensemble covariance between the 2-metre temperature and the soil moisture provides the needed sensitivity.

The LETKF suppresses the degrading effect of spurious correlations caused by the limited ensemble size with an independent, localised analysis for every grid point and a weighting of observations according to their distance from the analysis grid point.
We base the localisation on the Gaspari-Cohn covariance function \citep{gaspari1999} with a horizontal localisation radius of $15\,\text{km}$ and a vertical localisation radius of $0.7\,\text{m}$, where the observations are assigned to the surface, i.e. to $0\,\text{m}$ above ground.
These radii reflect typical error length-scales in the land surface and dampen the physically spurious impact of 2-metre temperature observations on soil layers below root depth.
We additionally use multiplicative prior covariance inflation with an inflation factor $\gamma = 1.006$ to avoid ensemble collapse (for details about localisation and inflation cf. \citealt{hunt2007}).

\subsection{Experiments}\label{experiments}

We define a single, deterministic run without data assimilation as our nature run, from which we also generate our observations (cf. \ref{sec:observations}).
We start this run with realistic and heterogeneous initial conditions at 2015-07-30 00:00 UTC and run it until 2015-08-07 18:00 UTC.
Further, we start four experiments as follows.

By perturbing the initial soil moisture saturation and soil temperature, we initiate an ensemble of 40 members, allowing 36 hours of omitted spin-up.
We perturb the initial soil fields with Gaussian noise, a zero mean, and a standard deviation of $0.06$ for the soil moisture saturation and of $1\,\text{K}$ for the soil temperature.
We correlate the noise with Gaussian kernels in horizontal and vertical direction, a scale of $14\,\text{km}$ or $0.5\,\text{m}$, and a truncation after $42\,\text{km}$ or $1\,\text{m}$, respectively.
Such correlations within the soil could be expected after a large-scale precipitation event.

Based around this ensemble of initial conditions, we run our two ensemble experiments.
An ensemble experiment without data assimilation, the open-loop run, is our reference.
With the spun-up ensemble conditions at 2015-07-31 12:00 UTC, we additionally start the localised ensemble Kalman filter experiment, where we hourly assimilate the sparsely-distributed 2-metre temperature observations into the soil moisture.

Using the ensemble mean of the initial conditions, we initialise a deterministic open-loop run without data assimilation.
Starting from its conditions at 2015-07-31 00:00 UTC, in the simplified extended Kalman filter experiment, we update the soil moisture once a day at 00:00 UTC with the interpolated 2-metre temperature field at 12:00 UTC.

\section{Results and Discussion}\label{sec:results}

\begin{figure}[ht]
    \centering
    \includegraphics[width=0.9\textwidth]{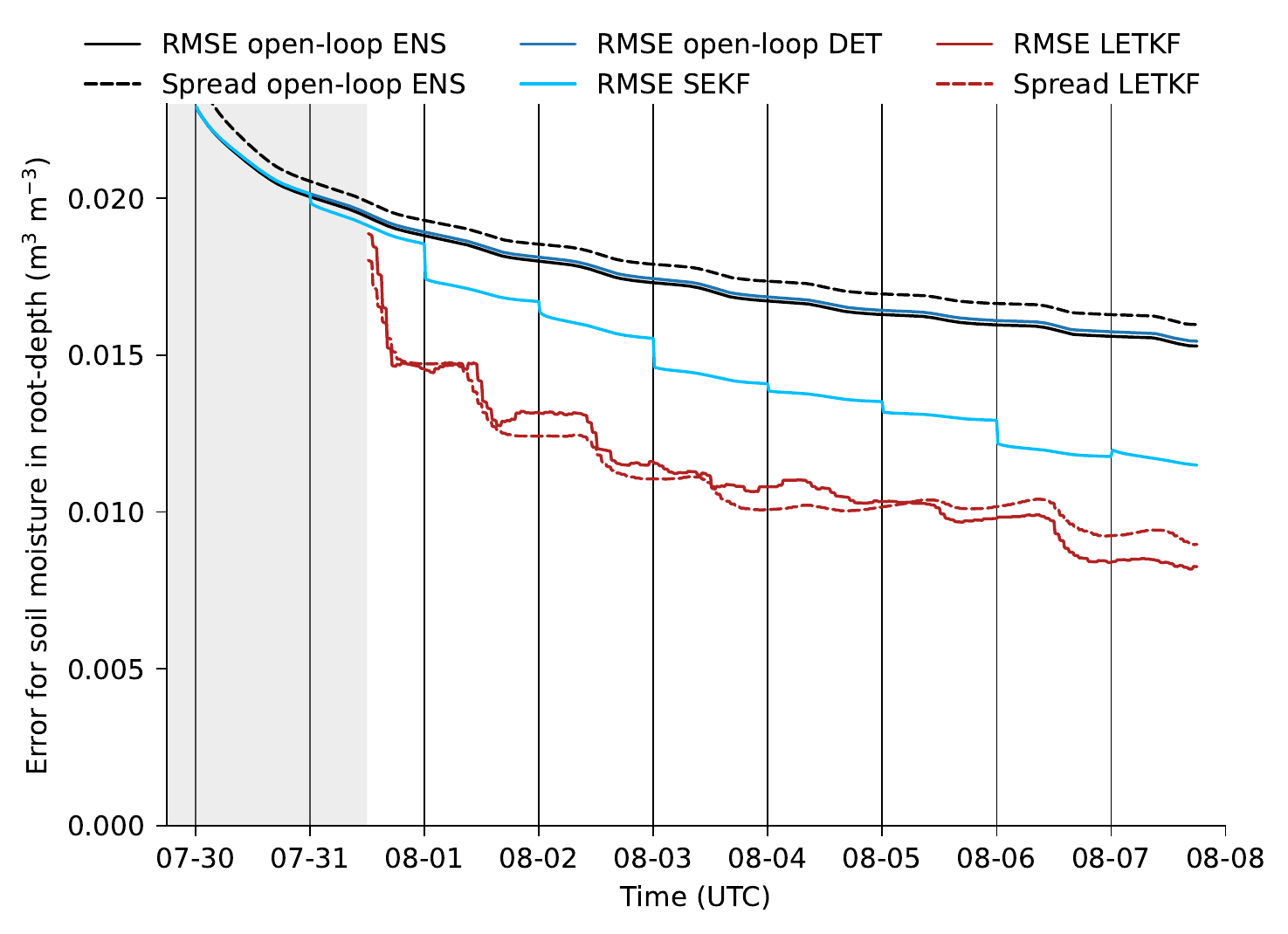}
    \caption{
        Spatial root-mean-squared error and ensemble standard deviation (spread) for the soil moisture in root-depth.\label{fig:h2o_rmse}
    }
\end{figure}

In the following, we will shortly describe and discuss our results.
We use the root-mean-squared error (RMSE) between our four experiments and the nature run as spatial-temporal mean (Table \ref{tab:exp_rmse}) and spatial mean (Figure \ref{fig:h2o_rmse}) to assess the performance of the temperature in 10 meters height and the soil moisture in root-depth.
For the ensemble-based experiments, we report the RMSE of the ensemble mean, whereas we also track the ensemble standard deviation.

\begin{table}[ht]
    \centering
    \caption{
        Root-mean-squared error and ensemble standard deviation (spread) for the temperature in 10 meters height (T10m) and soil moisture in root-depth (SM), estimated with the background forecasts on an hourly interval from 2015-07-31 13:00Z to 2017-08-07 18:00Z.\label{tab:exp_rmse}
    }
    \begin{tabular}{l|cc|cc}
        \headrow
        Experiment & RMSE T10m (K) & Spread T10m (K) & RMSE SM ($\text{m}^3 \text{m}^{-3}$) & Spread SM ($\text{m}^3 \text{m}^{-3}$)\\
        \hiderowcolors
        Open-loop ensemble & 0.158 & 0.172 & 0.0169 & 0.0173\\
        Open-loop deterministic & 0.178 & - & 0.0171 & - \\
        LETKF & 0.105 & 0.116 & 0.0114 & 0.0111\\
        SEKF & 0.118 & - & 0.0145 & -\\
    \end{tabular}
\end{table}

Initialised with the same initial conditions, the open-loop ensemble and the deterministic open-loop have similar forecast errors for the temperature and soil moisture (Table \ref{tab:exp_rmse}).
The different type of forecast results nevertheless in a smaller error for the ensemble than for the deterministic forecast, as should be expected.
For the atmospheric-boundary-layer temperature, both data assimilation methods reduce the forecast error to the same extent; their differences have the same order of magnitude as for the open-loop runs.
However, the LETKF has a larger impact on the soil moisture, where it improves the forecast by $33\%$ compared to $15\%$ with the SEKF.
Consequently, hourly updates with the LETKF are more efficient for the soil moisture analysis than daily updates with the SEKF.

Having the same lateral boundary conditions, all simulations converge towards the nature run for the soil moisture (Figure \ref{fig:h2o_rmse}).
Both the SEKF and LETKF improve their forecasts over time compared to the open-loop runs, showing a positive assimilation impact also after a few days of updating time.
The LETKF nevertheless reduces the initial departures between the forecast and nature run within its first update cycles much stronger than the SEKF.
Furthermore, the LETKF gains during daytime, whereas its updates have a neutral impact during night.
Thus, frequent, hourly, filtering updates with sparse observations have a positive impact compared to daily smoothing updates with a field.
This gain additionally suggests that the soil moisture could be included into the state vector of mesoscale data assimilation.

In a real-world setting, the SEKF needs an intermediate interpolation step.
This introduces an additional source of uncertainty, which can harm the assimilation.
As the LETKF can directly assimilate screen-level observations at their respective position, it can reduce uncertainties within the assimilation chain.

\begin{table}[ht]
    \centering
    \caption{
      The assimilation impact in $10^{-3}~\text{m}^3 \text{m}^{-3}$ for the soil moisture in root-depth.
      We define the impact as the difference in the spatio-temporal averaged RMSE for analyses with either a finite-differences or an ensemble approximation of the sensitivities compared to the background forecasts in the SEKF experiment.
      A positive number corresponds to an error reduction.
      \label{tab:impact}
    }
    \begin{tabular}{l|c}
        \headrow
        Approximation         & Impact\\
        \hiderowcolors
        Finite-differences    & 0.46\\
        Ensemble              & 1.13\\
    \end{tabular}
\end{table}

To see if the gain of the LETKF is only due to its more frequent updates, we evaluate the update step of the SEKF experiment (Table \ref{tab:impact}).
Here, by reusing the daily background forecasts of the SEKF experiment, we simply create daily analyses without restarting the model.
In the additional analyses, we approximate the sensitivities from the 2-metre temperature to the soil moisture in root-depth by the covariances from our open-loop ensemble.
This configuration resembles the current operational implementation at the ECMWF for their land surface data assimilation \citep{ecmwf_ifs_da_2020}, where the sensitivities are estimated based on covariances from their ensemble data assimilation system.

The assimilation impact with the ensemble approximation for the sensitivities is more than doubled compared to the finite-differences approximation (Table \ref{tab:impact}).
This additional error decrease indicates more accurate sensitivities from the ensemble approximation and supports the current implementation of the SEKF at the ECMWF.
Furthermore, the gain of the LETKF compared to the SEKF can be partially related to its ensemble approximation of the sensitivities.

\begin{figure}[ht]
\centering
\includegraphics[width=0.6\textwidth]{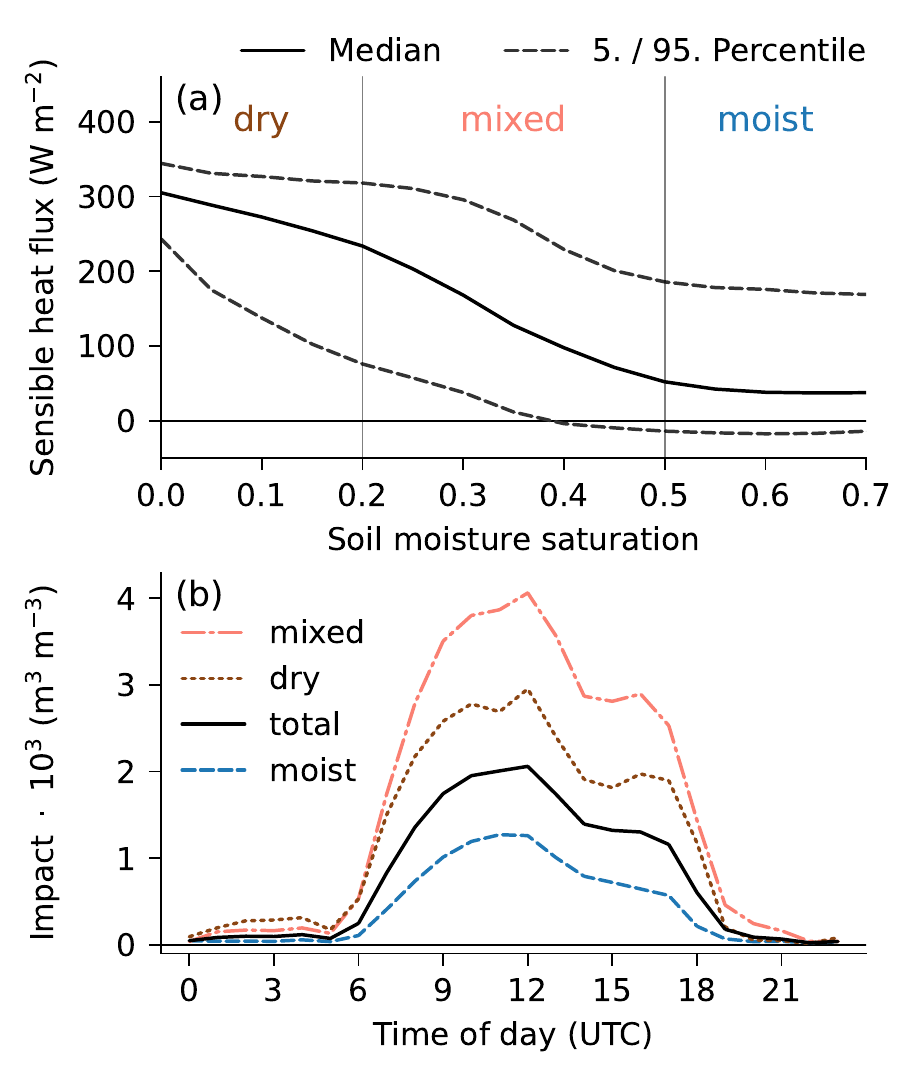}
    \caption{
        (a) The dependency of the sensible heat flux on the soil moisture saturation in the nature run for all grid points and days between 2015-07-31 to 2015-08-07 at 15:00 UTC.
        (b) Area mean diurnal cycles of the assimilation impact for the soil moisture in root-depth, valid from 2015-07-31 12:00 UTC to 2015-08-07 11:00 UTC.
        The soil moisture is classified with regimes determined in (a) on the basis of the nature run.
        We define the assimilation impact as spatio-temporal averaged RMSE difference between the ensemble mean of background forecasts in the LETKF experiment and analyses where we assimilate perfect-observed 2-metre temperature fields with a one-dimensional ensemble Kalman filter.
        A positive number corresponds again to an error reduction.\label{fig:temporal}
    }
\end{figure}

The sensible heat flux couples the atmospheric boundary layer temperature to the land surface temperature and drives changes in the 2-metre temperature during daytime.
This coupling process depends non-linearly on the soil moisture saturation (Figure \ref{fig:temporal}, a); the 2-metre temperature accordingly exhibits a similar non-linear dependency.
For moist soils (saturation $> 0.5$), the evapotranspiration is unconstrained from the available water, and the sensible heat flux is insensitive to small changes in the soil moisture.
Near the wilting point (saturation $< 0.2$), in dry soils, the available water suppresses transpiration, and the sensible heat flux shows only little sensitivity to the soil moisture.
As a consequence, grid points with mixed soil moisture regimes have the highest sensitivity.
There, we also expect the highest assimilation impact from 2-metre temperature observations for the soil moisture.

As the number of observations limits the assimilation, we again perform analyses without restarting the model, this time based on our hourly LETKF background forecasts.
Using a one-dimensional ensemble Kalman filter, which takes only vertical sensitivities between 2-metre temperature and soil moisture into account, we assimilate interpolated 2-metre temperature fields without any observational error into the soil moisture.

Because the coupling strength between land surface and atmospheric boundary layer drives the diurnal cycle of the assimilation impact (Figure \ref{fig:temporal}, b), its highest values are around noon with almost no impact during night.
The modulation of the diurnal cycle by the temporal development of the atmospheric boundary layer results into a small reinforcement of the assimilation impact around 17:00 UTC, which needs further investigation.
In accordance to the non-linear dependency of the sensible heat flux upon the soil moisture, we find the highest impact for grid points with mixed soil moisture regimes and the lowest for moist grid points.
Hence, the soil moisture itself is an influencing factor on the assimilation, and two-metre-temperature observations above mixed soil moisture saturations have the highest information content for the soil moisture in root-depth.

Above very dry and humid soils, the global non-linear coupling structure constrains the information content.
As information from the land surface are propagated with a delay into the atmospheric boundary layer, four-dimensional ensemble data assimilation methods like the 4D-LETKF \citep{harlim2007, kalnay2007} or iterative ensemble Kalman methods \citep{kalnay2010, bocquet2014} are an opportunity to take this delay into account.
In a complementary approach, we can allow for non-linear structures in the data assimilation by assimilating features of observations in a feature-based data assimilation framework \citep{morzfeld2018}.
These methods can help to extract more information about the soil moisture from screen-level observations.

\begin{figure}[ht]
    \centering
    \includegraphics[width=0.6\textwidth]{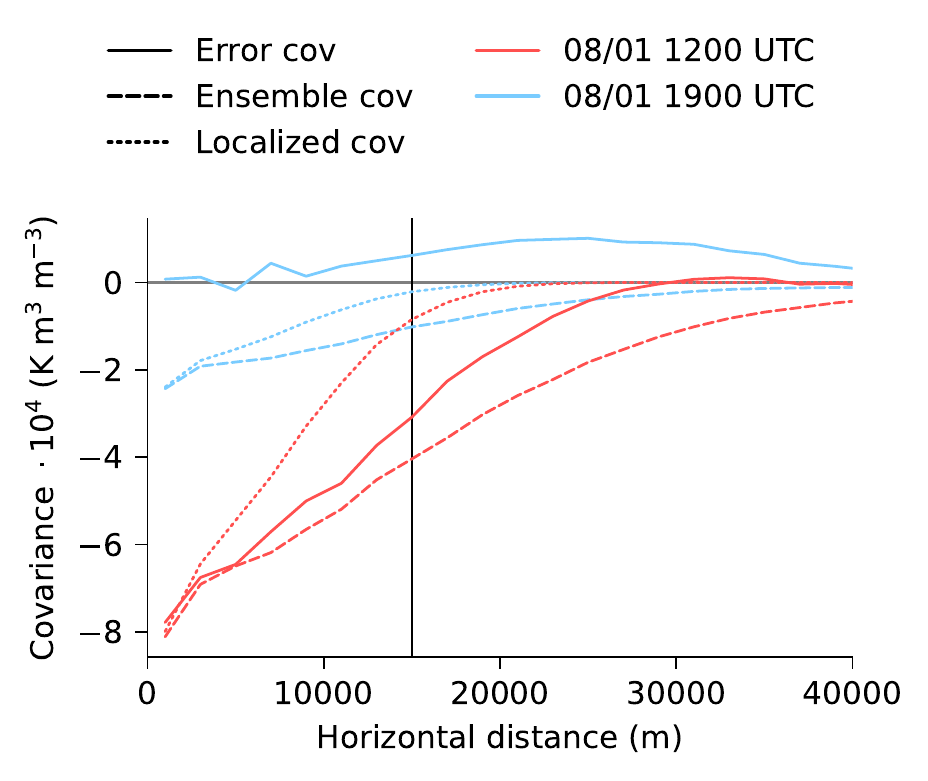}
    \caption{
        Covariances between 2-metre temperature and soil moisture in root-depth as function of the horizontal distance based on randomly-sampled $2\times 10^6$ grid point pairs
        The error covariances are the binned averaged error covariances of the ensemble mean.
        The vertical line at $15\,\text{km}$ represents the chosen localisation radius.\label{fig:localisation}}
\end{figure}

As shown in Figure \ref{fig:h2o_rmse}, we can directly assimilate 2-metre temperature observations into the soil moisture by making use of horizontal and vertical covariances within an ensemble Kalman filter.
These ensemble covariances should match in the ideal case the error covariances of the ensemble mean to the nature run.
Although exhibiting a too long correlation radius, the ensemble covariances can reproduce the error covariances around noon (Figure \ref{fig:localisation}).
For the shown evening case, the ensemble is unable to represent the changed covariance conditions from the nature run.
This exposes possible issues with the ensemble covariances in transition periods from day to night.
There, we expect a negative assimilation impact, and the best would be a deactivated assimilation.
Localisation with our chosen radius of $15\,\text{km}$ regularises the ensemble covariances too much during daytime and too little during night.
In such settings, we can likely improve the analysis by using a dynamic and daytime-dependent localisation radius.

\section{Conclusion}\label{conslusion}

In this study, we assimilate 2-metre temperature observations into the soil moisture of a coupled limited-area atmosphere-land modelling system for a seven-day summer period.
Based on the results in our idealised twin experiments, we conclude the following:

\begin{itemize}
    \item
        A localised three-dimensional ensemble Kalman filter has an emerging potential to improve the soil moisture analysis compared to a one-dimensional simplified extended Kalman filter.
        Although both approaches have almost the same assimilation impact onto the temperature in the atmospheric boundary layer, we find an error reduction for the soil moisture with a three-dimensional ensemble Kalman filter.
        Moreover, we can assimilate with the ensemble Kalman filter 2-metre temperature observations into the soil moisture on an hourly basis, in an approach similarly used for operational mesoscale data assimilation.
    \item
        Replacing the finite-differences approximation to the sensitivities with an external ensemble approximation, we achieve a higher assimilation impact in the update step of a simplified extended Kalman filter.
        This supports the current implementation at the ECMWF \citep{ecmwf_ifs_da_2020}, where the sensitivities are approximated with runs from their ensemble data assimilation system.
        Thus, we attribute the advantage of the three-dimensional ensemble Kalman filter to its ensemble approximation of the sensitivities and its more frequent updates.
    \item
        The hourly updates in an ensemble Kalman filter lead to a positive impact during daytime and a neutral impact during night.
        The soil moisture itself modulates hereby the information content of screen-level observations for the soil moisture.
        The non-linear dependency of the sensible heat flux upon the soil moisture constrains the sensitivity of the 2-metre temperature to changes in the soil moisture.
        Consequently, we confirm the highest sensitivity and also the largest assimilation impact for soil conditions with a soil moisture saturation between $0.2$ and $0.5$.
    \item
        By using localisation and diagonal covariances in an ensemble Kalman filter, we can directly assimilate screen-level observations at their respective position into the soil moisture.
        This allows us to skip the intermediate interpolation step, needed in the simplified extended Kalman filter, and to reduce observational uncertainties.
        We additionally discover a daytime dependency of the errors in our vertical ensemble covariances, and the analysis could be improved with a dynamic localisation radius.
        As demonstrated by our assimilation gain, we can nevertheless directly assimilate screen-level observations with one single localisation radius into the land surface.
\end{itemize}

\section*{Acknowledgements}
    This work is a contribution to the project FOR2131, “Data Assimilation for Improved Characterization of Fluxes across Compartmental Interfaces”, funded by the “Deutsche Forschungsgemeinschaft” (DFG, German Research Foundation) under grant 243358811.
    We would thank the whole research unit FOR2131 for discussions along the track and especially Bernd Schalge for providing us the data for the initial spin-up and lateral boundary conditions.
    We additionally want to acknowledge researchers at the “Deutscher Wetterdienst” for helpful hints in the initial phase of the research.
    At last but not least, we want to appreciate Marc Bocquet and three anonymous reviewers for providing helpful insights, which have improved the manuscript.

\section*{Conflict of interest}
    The authors declare that they have no conflict of interest.

\bibliography{22_soil_assimilation.bib}

\end{document}